\documentclass[3p,times,procedia]{elsarticle}
\usepackage{nupha_ecrc}

\volume{00}

\firstpage{1}

\journalname{Nuclear Physics A}

\runauth{}

\jid{nupha}

\jnltitlelogo{Nuclear Physics A}

\usepackage{amssymb}

\usepackage{lineno}

\usepackage[figuresright]{rotating}

\usepackage{booktabs}
\usepackage{upgreek}
\usepackage{multirow}

\begin{document}

\begin{frontmatter}



\dochead{XXVIIth International Conference on Ultrarelativistic Nucleus-Nucleus Collisions\\ (Quark Matter 2018)}

\title{Upgrade of the ALICE central barrel tracking detectors: ITS and TPC}


\author{P. Gasik on behalf of the ALICE Collaboration}

\address{Physik Department E62, Technische Universit\"{a}t M\"{u}nchen, Garching, Germany\\and\\Excellence Cluster 'Origin and Structure of the Universe', Garching, Germany}

\begin{abstract}
The ALICE Collaboration will undertake a major upgrade of the detector apparatus during the second LHC Long Shutdown LS2 (2019-2020) in view of the Runs 3 and 4 (2021-2029). The objective of the upgrade is two-fold: i) an improvement of the tracking precision and efficiency, in particular in the low-momentum range; ii) an improvement of the readout capabilities of the experiment, in order to fully exploit the luminosity for heavy ions envisaged after LS2.
The first goal will be achieved by replacing the Inner Tracking System with a new tracker, composed of seven layers of silicon pixel detectors. The new tracker will be made up of about 25000 Monolithic Active Pixel Sensors with fast readout, resulting in a material thickness reduced to 0.3\% (inner layers) -- 1\% (outer layers) of the radiation length and a granularity of $28\times28$\,$\upmu$m$^2$. The second goal will be achieved, among other measures, by replacing the readout chambers of the 90\,m$^3$ Time Projection Chamber with Micro Pattern Gaseous Detectors. In particular, the new readout chambers will consist of stacks of 4 Gas Electron Multiplier foils combining different hole pitches. The upgraded detector will operate continuously without the use of a triggered gating grid. It will thus be able to record all Pb--Pb collisions at the LHC interaction rate of 50\,kHz. 
\end{abstract}


\end{frontmatter}


\section{Introduction}
\label{pgasik:sec:intro}
After the second Long Shutdown (LS2, 2019-2020) the LHC will deliver Pb beams colliding at an interaction rate of about 50 kHz. This will result in a significant improvement on the sensitivity to rare probes that are considered key observables to characterise the QCD matter created in such
collisions. The only way to fully exploit the LHC potential in Runs 3 and 4 (2021-2029) for both low momentum and high momentum probes is to record all Minimum Bias (MB) events delivered by the collider. The ALICE Collaboration \cite{ALICE} aims at integrating a luminosity of 13\,nb$^{-1}$ Pb--Pb collisions, which represents a MB data sample larger by a factor 50-100 with respect to the ongoing Run 2 \cite{LOI}. 

In addition to the enhanced readout capabilities, a significant improvement of the tracking precision and efficiency, in particular in the low-momentum range, will be reached by increasing the tracking granularity, reducing the material budget close to the interaction point (IP), and minimising the distance of the first measurement layers to the IP.

These goals will be achieved by replacing the Inner Tracking System (ITS) with a new tracker based on Monolithic Active Pixel Sensor (MAPS) technology and upgrading the ALICE Time Projection Chamber (TPC) with Gas Electron Multiplier (GEM) \cite{Sauli} based readout chambers \cite{LOI}. The status of both projects will be discussed in the following contribution.
\section{Inner Tracking System Upgrade}
\label{pgasik:sec:its}
The ALICE Inner Tacking System (ITS) will be replaced by a new, light-weight, high-resolution apparatus fully based on the Monolithic Active Pixel Sensor (MAPS) technology \cite{ITSTDR}. The new MAPS chip ALPIDE (ALice PIxel DEtector) was developed in the 180\,nm CMOS TowerJazz process \cite{TJ}, where the circuits are fabricated on substrates with a 25\,$\upmu$m thick, high-resistivity epitaxial layer on highly doped P-type substrate. The electrons released in the epitaxial layer can diffuse laterally while remaining vertically confined by potential barriers at the interfaces between the epitaxial layer with the overlying P-wells and the underlying P-type substrate. Thanks to that, all primary electrons will reach the depletion zone created at the junction of the N-well sensing diode and epitaxial layer giving almost a 100\% detection efficiency. The sensing diode is 2$\times$2\,$\upmu$m$^2$ large which is a factor of $\sim$100 lower than the pixel size. This, together with a large depletion zone, results in a very small input capacitance and thus relatively high amplitude of the induced signals ($\sim$40\,mV for MIPs). The RMS electron noise is $\sim$5$e^-$ resulting in a large signal-to-noise ratio \cite{ITSTDR, ALPIDE, ALPIDE2}. 

The ALPIDE chip is 15$\times$30\,mm$^2$ large and contains more than half a million pixels organised in 1024 columns and 512 rows. Its very low power consumption ($<$40\,mW/cm$^2$), a fake hit rate of less than 10$^{-9}$ per pixel per event and spatial resolution of 5\,$\upmu$m make it a perfect choice for the inner tracker of the ALICE experiment. Table~\ref{tab:its} shows a comparison of several key parameters between the old and new detectors \cite{ITSTDR, ALPIDE, ALPIDE2}.
\begin{table*}[htp]
\caption[]{Comparison between old and new ITS (inner layers) parameters.}
\label{tab:its}
\centering
\begin{tabular}{lccc}
\toprule
&	Old ITS		&	New ITS \\\midrule
Distance to the IP & 39\,mm & 22\,mm\\
Material budget ($X_0$) & 1.14\% & 0.30\%\\
Pixel size & 50$\times$425\,$\upmu$m$^2$ & 27$\times$29\,$\upmu$m$^2$\\
Granularity & 20 channels/cm$^3$ & 2000 pixels/cm$^3$\\
\multirow{2}*{Readout rate} & \multirow{2}*{1\,kHz} & 100\,kHz (Pb--Pb)\\
 && 1\,MHz (pp)\\
\bottomrule
\end{tabular}
\end{table*}

The active pixel area of the new ITS will cover almost 10\,m$^2$ and consist of seven cylindrical layers (3 Inner Barrel + 4 Outer Barrel) of ALPIDE chips summing up to $\sim$12.5$\times$10$^9$ MAPS \cite{ITSTDR}. The pixel chips are installed on staves from 22\,mm to 400\,mm around the interaction point. An improvement of the impact parameter resolution by a factor of three in the transverse plane and by a factor of five along the beam axis is expected, reaching for example 20\,$\upmu$m in both directions at a transverse momentum of 1 GeV/$c$. The overall tracking efficiency and the $p_{\mathrm{T}}$ resolution at low $p_{\mathrm{T}}$ will improve as well \cite{ITSTDR}.
\begin{figure}
\begin{center}
\includegraphics[width=0.74\linewidth]{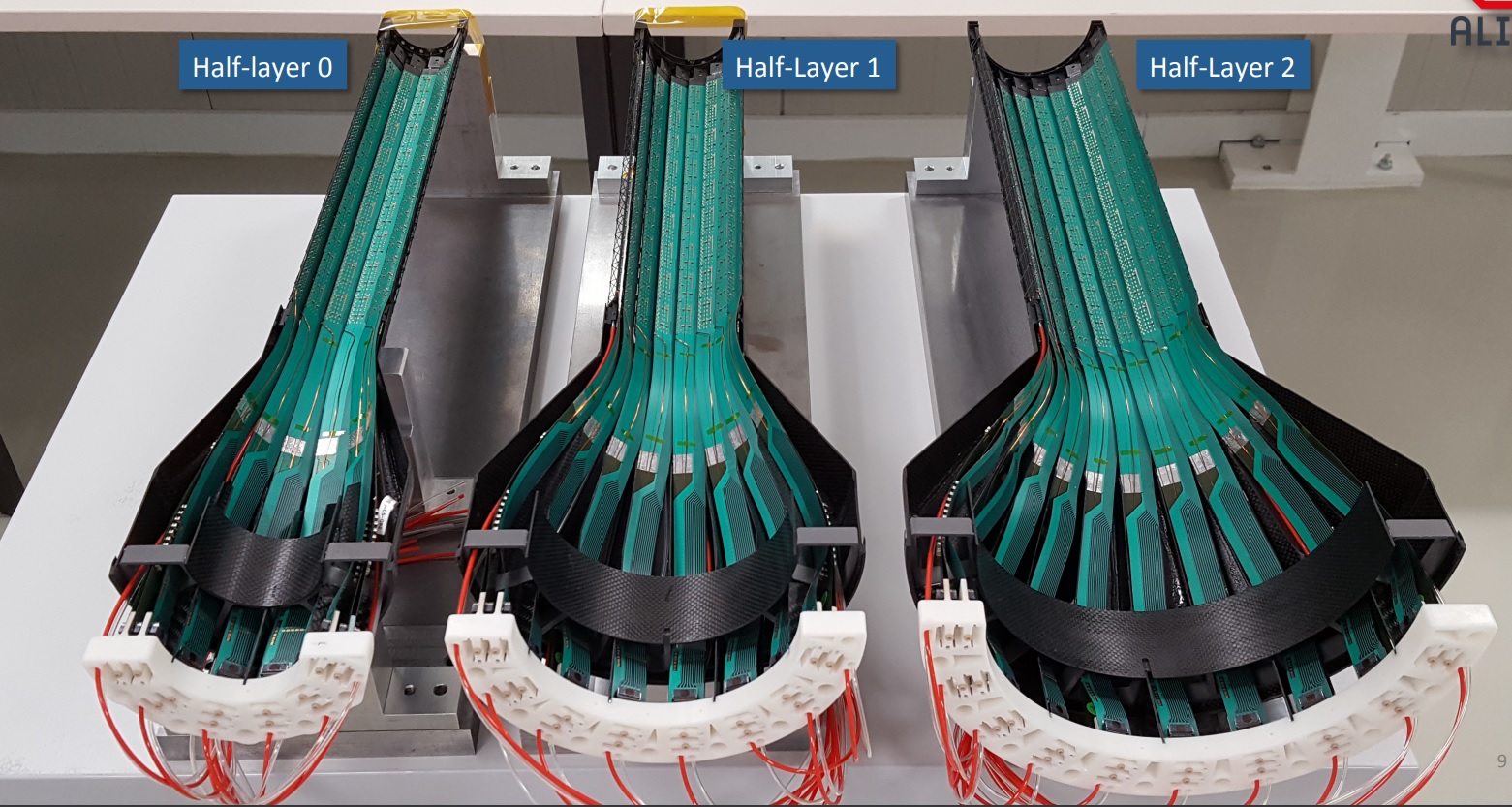}
\caption{Photograph of the upgraded ITS inner half-layers}
\label{fig:its}
\end{center}
\end{figure}

At the time of submission of this manuscript, the CMOS manufacturing is completed. The assembly and testing of the ALPIDE modules (CERN, Bari, Liverpool, Pusan, Strasbourg, Trieste, Wuhan) is planned until September 2018. The stave production (Berkeley, CERN, Daresbury, Frascati, NIKHEF, Torino) will continue until April 2019. The first three half-layers of the Inner Barrel have been recently finalised (see Fig.~\ref{fig:its}). The construction of all mechanical structures (Berkeley, CERN, Padua) has been completed.
After the assembly, the ITS will be commissioned on the surface followed by the installation in the ALICE cavern in 2020.
\section{Time Projection Chamber Upgrade}
\label{pgasik:sec:tpc}
A large Time Projection Chamber is the main device for tracking and charged
particle identification in the ALICE experiment \cite{TPC}. At the interaction rates envisaged after the LHC LS2, an average pile-up of about five events is expected in the drift volume of the TPC. To overcome the rate limitations imposed by the present gated readout scheme, the existing Multi-Wire Proportional Chambers will be replaced by continuously operated GEM detectors \cite{TDR}.

The ungated operation of the ALICE TPC will lead to a considerable accumulation of positive ions in the drift volume that emerge from the amplification region. The resulting space-charge distortions must be kept sufficiently low to allow for efficient online track reconstruction and distortion corrections. To fulfil the challenging requirements of the upgrade, a novel configuration of GEM detectors has been developed which maintains the excellent particle identification of the current TPC, and provides efficient ion trapping and operational stability by stacking four GEM foils operated at a specific field configuration \cite{TDR, Dedx, Addendum, mpgd17}. Quadruple GEM (4-GEM) readout chambers will employ stacks containing Standard (S, 140\,$\upmu$m pitch) and Large Pitch (LP, 280\,$\upmu$m pitch) GEMs, in the configuration S-LP-LP-S.

Forty Inner and forty Outer ReadOut Chambers (IROC and OROC) will be built (including spares) to equip the TPC in 2019, when the entire detector will be moved to the surface for the upgrade. The production of the chambers is divided into several steps and shared between many institutes in Europe and in the US. It includes: $i)$ GEM foil production and quality assurance (CERN, Helsinki, Budapest), $ii)$ GEM framing onto 2\,mm thick G11 frames (Wayne State University, TU Munich, Univ. Bonn, GSI), $iii)$ chamber body assembly (Knoxville, Univ. Frankfurt, Univ. Heidelberg), $iv)$ readout chamber assembly and commissioning (Yale, GSI, HPD Bucharest) and $v)$ the final acceptance, storage and installation (CERN). Details of various production steps can be found in \cite{mpgd17AM} and references therein. In order to track all ROC components, shipments and test results, a custom-made database has been developed.

\begin{figure}
\begin{center}
\includegraphics[width=0.74\linewidth]{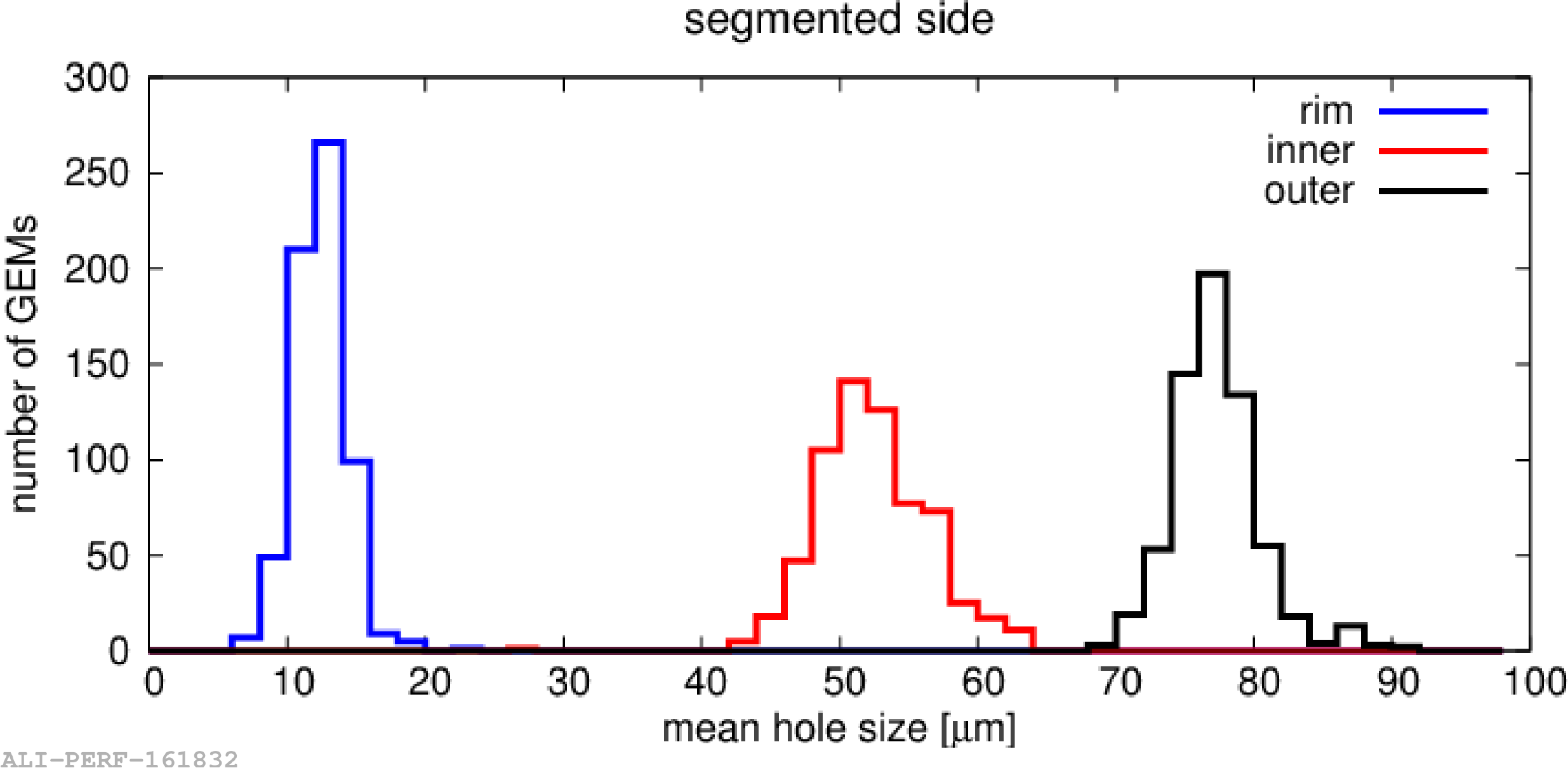}
\caption{Distribution of GEM hole sizes of 647 GEMs produced for the ALICE TPC Upgrade}
\label{fig:GEM}
\end{center}
\end{figure}

In total, 640 GEMs will be used to equip the chambers with quadruple GEM stacks (one stack per IROC and three stacks per OROC). After the production in the CERN PCB workshop, the GEMs are shipped to the advanced quality assurance site. Each GEM hole is photographed and its inner and outer diameters are measured. The scan allows to produce hole size uniformity and defect maps which are the base for the detailed assessment of the foil quality.
Figure~\ref{fig:GEM} shows an exemplary distribution of the inner and outer hole diameter, and the rim sizes for 647 GEMs. The mean values of 12.39\,$\upmu$m, 52.39\,$\upmu$m and 77.17\,$\upmu$m for the rim, inner and outer diameter, respectively, are well in agreement with the design values.

After the advanced quality assurance, the foils are first shipped to the framing sites and then to the ROC assembly sites. The assembled chambers are commissioned in a number of tests, including a gas tightness test, gain curve and energy resolution measurement, gain and ion backflow uniformity scans,
 and finally the full X-ray irradiation considered as a stress test in which current densities of 10\,nA/cm$^2$ at a gain of 2000 are induced \cite{mpgd17AM}. Part of the production chambers are also tested at the LHC, in the ALICE cavern, few metres from the interaction point in the forward pseudorapidity region. Full high voltage is applied to the GEMs while being irradiated by the collision products. The load on the chambers is comparable to the load expected in their nominal position in Runs 3 and 4. Up to now, 10 chambers were successfully tested in this environment.

The chamber production will last until October 2018 and their installation is planned for Spring 2019.
\section{Summary and Outlook}
\label{}
About 10-fold increase of Pb--Pb delivered luminosity is expected in Runs 3 and 4. The ALICE experiment will undergo a major upgrade in order to record all MB events at 50\,kHz Pb--Pb collisions which is a factor 50-100 more than in Run 2. The new Inner Tracking System based on the ALPIDE MAPS chips will be installed to enhance the tacking and vertexing performance. The upgraded TPC with GEM-based readout chambers will be read out continuously preserving the present tracking and PID capabilities. 
\section*{Acknowledgements}
The speaker acknowledges support by the DFG Cluster of Excellence "Origin and Structure
of the Universe" (www.universecluster.de) [project number DFG EXC153], the Federal
Ministry of Education and Research (BMBF, Germany) [grant number 05P15WOCA1] and The Helmholtz Graduate School for Hadron and Ion Research "HGS-HIRe for FAIR".





\bibliographystyle{elsarticle-num}
\bibliography{<your-bib-database>}



\end{document}